\documentclass[aps,pre,twocolumn,floatfix,showpacs,superscriptaddress]{revtex4} 
\usepackage{graphicx}
\usepackage{epsfig}
\usepackage{amssymb,amsmath}
\usepackage{bm}

\begin{document}
\title{Layering, freezing and re-entrant melting of hard spheres in soft confinement}

\author{Tine Curk}
\affiliation{University of Cambridge, Lensfield Road, Cambridge CB2 1EW, United Kingdom }
\author{Anouk de Hoogh}
\affiliation{FOM Institute AMOLF, Science Park 104, 1098 XG, Amsterdam, The Netherlands }
\author{Francisco J. Martinez-Veracoechea}
\affiliation{University of Cambridge, Lensfield Road, Cambridge CB2 1EW, United Kingdom }
\author{Erika Eiser}
\affiliation{University of Cambridge, Cavendish Laboratory, J.J. Thomson Avenue,
Cambridge CB3 0HE, United Kingdom}
\author{Daan Frenkel}
\affiliation{University of Cambridge, Lensfield Road, Cambridge CB2 1EW, United Kingdom }
\author{Jure Dobnikar}
\email[]{jd489@cam.ac.uk}
\affiliation{University of Cambridge, Lensfield Road, Cambridge CB2 1EW, United Kingdom }
\affiliation{Institute Jo\v zef Stefan, Jamova 39, SI-1000, Ljubljana, Slovenia }
\author{Mirjam E. Leunissen}
\email[]{m.e.leunissen@amolf.nl}
\homepage[]{www.amolf.nl}
\affiliation{FOM Institute AMOLF, Science Park 104, 1098 XG, Amsterdam, The Netherlands }

\date{\today}

\begin{abstract}
Confinement can have a dramatic effect on the behavior of all sorts of
particulate systems and it therefore is an important phenomenon in
many different areas of physics and technology. Here, we investigate
the role played by the softness of the confining potential. Using
grand canonical Monte Carlo simulations, we determine the phase
diagram of three-dimensional hard spheres that in one dimension are constrained to a plane by a harmonic potential.
The phase behavior depends strongly on the density and on the stiffness of the harmonic confinement.
Whilst we find the familiar sequence of confined hexagonal and square-symmetric packings, we do not observe any of the
usual intervening ordered phases. Instead, the system phase separates
under strong confinement, or forms a layered re-entrant liquid phase
under weaker confinement. It is plausible that this behavior is due to the larger
positional freedom in a  soft confining potential and to the contribution that the confinement energy makes to the total free energy. The fact that specific structures can be
induced or suppressed by simply changing the confinement conditions
(e.g. in a dielectrophoretic trap) is important for applications that
involve self-assembled structures of colloidal particles.
\end{abstract}


\maketitle

\section{Introduction}
The behavior of particles in confined geometries is important in many
different areas of physics and technology. This includes the physics
of ions in electromagnetic traps \cite{walz}, of dusty plasmas
confined by external fields \cite{totsuji,messina}, of classical
electrons in quantum wells \cite{esfarjani} and of colloidal
suspensions in narrow slits \cite{lowen}, as well as
application-oriented topics in nanotechnology, (bio)lubrication and
the self-assembly of microstructured materials (e.g. Ref. \cite{krishnan,leckband,copley,cohen,vlasov}). It is well-known that
confinement effects can dramatically change the behavior of such
systems, both quantitatively and qualitatively. For instance, when a
suspension of colloidal hard spheres is confined in a wedge-shaped
geometry, one observes a  rich cascade of different
equilibrium crystal structures as the wall spacing increases
\cite{pieranski,pansu,neser,fortini}. Such behavior contrasts sharply 
with the
bulk phase diagram, which consists of a single, density-dependent
liquid to face-centered-cubic crystal transition. The past few decades
have seen many studies of the behavior of hard spheres between
impenetrable walls (e.g. Ref. \cite{fortini,schmidt,schmidt2,neser}), of
charged particles under strong confinement (a model for trapped
Coulombic or Yukawa particles, e.g. Ref. \cite{totsuji,messina,dubin,fontecha}), and of confined dipolar colloids 
(e.g. Ref. \cite{maret,Osterman,Dobnikar}). However, to the best of our knowledge,
there have been no studies that systematically investigate the role
played by the softness of the confining potential. This is an
important issue to address, as more and more systems become available
that involve some form of soft confinement. In the areas of colloid
science and nanotechnology one can for instance think of suspensions
confined in dielectrophoretic \cite{docoslis,sullivan,leunissen} or
laser-optical fields \cite{ackerson,jenkins}, nanometric objects in charged
slits \cite{krishnan}, particles interacting with soft polymer
substrates \cite{akcora,eiser}, or particles trapped at liquid-liquid
or liquid-gas interfaces \cite{Binks,Clegg,leunissen2,oettel,delGado}. Using Monte Carlo
simulations, we here study three-dimensional systems of hard spheres that in one dimension are constrained to a plane by a
harmonic potential. Unlike hard boundaries, this soft potential well
does not prescribe a particular confinement `width', can be
continuously tuned from very strong to very weak confinement, and makes an energy contribution to the total free energy which depends on the exact particle positions.
We
highlight the unique properties imparted by such soft confinement by
comparing the observed phase behavior with that of hard spheres
between two hard walls \cite{fortini}, as well as with the behavior of
a more complex system of highly charged particles and their
counterions between neutral walls \cite{oguz}. In the latter system,
the particles experience a combination of an effective harmonic
potential due to the counterions and long-ranged repulsive Coulomb
interactions.

\section{Model}
We performed grand canonical Monte Carlo simulations of hard spheres that are constrained to a plane by a harmonic potential. Hard spheres do not interact with
each other unless their cores overlap and for any pair of particles
the interaction potential is given by:
\begin{eqnarray}
\frac{U_{\rm{sphere-sphere}}}{k_BT}(r) = \left\{
\begin{array}{ll} 0,
& r\geqslant \sigma \nonumber\\
\infty, & r < \sigma \end{array} \right.
\end{eqnarray}
where $k_B$ is the Boltzmann constant, $T$ is the absolute
temperature, $r$ is the center-to-center distance of the particles and
$\sigma$ is the particle diameter, which we took as the unit of length
in our simulations. We used periodic boundary conditions in the $xy$
plane (box size $L_x \times L_y$ = $20 \times$ 20, unless stated
otherwise) and a soft confining harmonic potential centered around $z
= 0$, which acts on each particle individually and whose softness was
set through the spring constant $k$ (in units of $\sigma^{-2}$):
\begin{equation}
\frac{U_{\rm{conf}}}{k_BT}(z) = \frac{kz^2}{2}
\end{equation}
We performed simulations for different chemical potentials of the
reservoir, starting with an empty box. Using cell lists and an early
rejection scheme \cite{Frenkel}, we typically performed $6\times
10^{10}$ Monte Carlo moves in which we attempted to insert, delete or
displace ($d_{\rm{max}} = 0.05$) a randomly chosen particle, where the
fraction of insertion and deletion moves was fixed at 0.2. We also
performed simulations with an additional move in which the $L_x/L_y$
aspect ratio of the box was allowed to change while keeping the area
$A=L_xL_y$ constant. In principle, this should prevent the box shape
from dictating the structure of the particle packing, but in all cases
the resulting structures were identical to those observed in a square
box.  

After thorough equilibration, we determined the number of
particles $N$ in the simulation box and from this the density, $\rho =
N/A$, and characterized the structure of the typically layered
particle packings by calculating the four-fold ($q_4$) and six-fold
($q_6$) symmetric two-dimensional bond order parameters in each of the
layers \cite{q}. The bond-order parameters consider all of the
nearest neighbors of a given particle that lie approximately in the
same $z$ plane. Based on geometric arguments and the observed particle distributions along the $z$ axis, the nearest neighbors were here defined as those particles
residing within a center-to-center distance $r_{\rm{NN}} \leqslant
1.3$ of the particle of interest and with a height difference $\delta
z_{\rm{NN}} \leqslant 0.3$. This definition excludes second-nearest neighbors, which in a close-packed square-symmetric layer reside at $r=\sqrt{2}$, and is also found to reliably discriminate between thermally broadened and adjacent layers along $z$. The ratio between the two bond order
parameters allowed us to distinguish between the three main types of
structures found in our simulations: disordered liquid for $1/3
\leqslant q_4/q_6 \leqslant 3$, hexagonally ($\triangle$) packed for
$q_4/q_6< 1/3$, and square ($\square$) packed for $q_4/q_6 > 3$. The crossover values were determined from plots of $q_4/q_6$ across the entire density range studied in the simulations, as well as the density probability distributions, which revealed the first-order phase transitions. We find that varying $r_{\rm{NN}}$ and $\delta z_{\rm{NN}}$ by $\pm0.1$ does not affect the resulting phase diagram.

\section{Results}

\begin{figure}[b]
\includegraphics[width=0.45\textwidth]{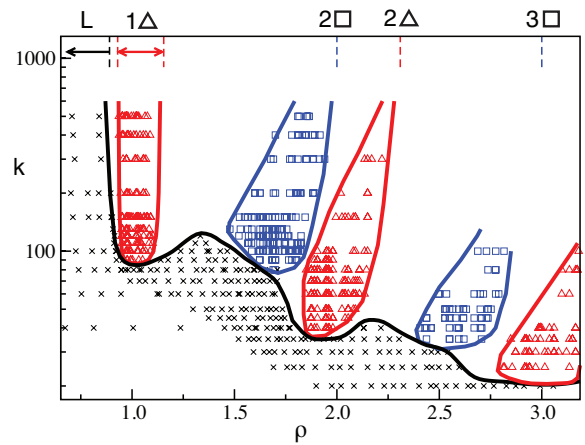}
\caption{\label{fig1} (Color Online) Phase diagram in the spring
  constant -- density representation. Every point represents the result
  of a single simulation run. Black crosses ($\times$): liquid. Red
  triangles ($\triangle$): hexagonally packed. Blue squares
  ($\square$): square packed (structures shown in Fig. \ref{fig2}). Approximate phase boundaries are
  indicated with solid lines. Dashed lines indicate the stable phases
  in the limit $k \to \infty$ (from Fig. \ref{fig3}).}
\end{figure}

\begin{figure}[]
\vskip 2mm
\includegraphics[width=0.45\textwidth]{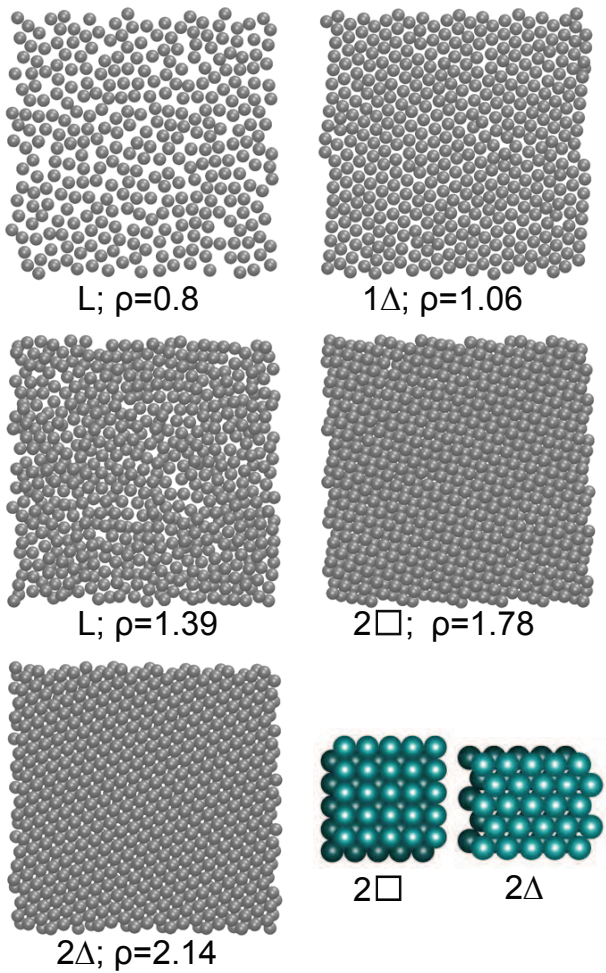}
\caption{\label{fig2} (Color Online) Simulation snapshots at different densities (for $k=100$) and `ideal' schematic representations (lower right) of the spatial arrangement of the particles in the square ($\square$) and hexagonally ($\triangle$) packed structures.}
\end{figure}

\begin{figure}[t]
\vskip 2mm
\includegraphics[width=0.45\textwidth]{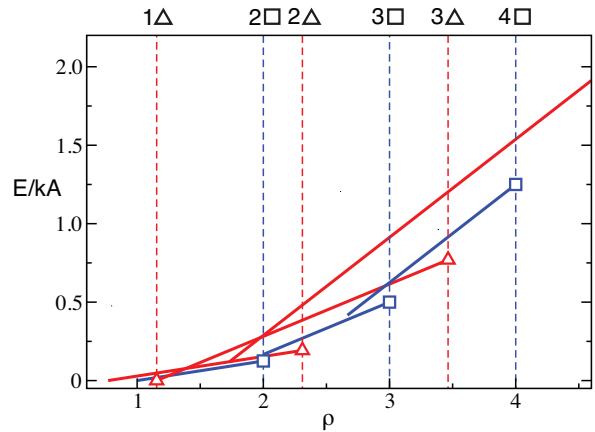}
\caption{\label{fig3} (Color Online) Normalized energies per unit area of the
  hexagonal (red lines, $\triangle$) and square (blue lines,
  $\square$) packed structures in the limit $k\to\infty$.}
\end{figure}

The phase diagram in Fig. \ref{fig1} and the simulation snapshots in Fig. \ref{fig2}
provide an overview of the
observed particle packings as a function of the density and spring
constant. At high spring constants ($k\gtrsim 150$), i.e. relatively
strong confinement, and low densities the particles form a layer of
disordered liquid ($L$) at the minimum of the confining harmonic
potential. At higher densities, the liquid freezes into one or more
crystalline layers which have a hexagonal ($\triangle$) or square
($\square$) symmetry (Fig. \ref{fig2}), following an alternating sequence as the density
increases: $1\triangle\to 2\square\to 2\triangle\to 3\square\to
3\triangle\to\ldots$ (the integers indicate the number of particle
layers that is formed). To rationalize these observations, we consider
the limit $k\to\infty$ (or $T\to 0$) in which the entropic
contribution to the free energy can be neglected and the energy contribution of the confining potential dominates:
$F\approx E = \frac{k}{2}\sum z^2$, where the sum runs over all the
particles in the system. Simple geometry then gives the energies of
the perfect hexagonal and square packed structures, which scale linearly
with the density because of the quadratic form of the confining
potential, Fig. \ref{fig3}. The maximum density of each of the phases
corresponds to close packing, e.g. $\rho_{\rm{max}}=2$ for $2\square$ and
$\rho_{\rm{max}}=4/\sqrt{3}$ for $2\triangle$. Starting at low density, the
$L\to 1\triangle$ transition is given by the two-dimensional hard disk
freezing transition \cite{Velasco,Binder} and the $1\triangle$ phase
(strongly constrained to $z = 0$ and with negligible energy) is
stable until close packing at $\rho=2/\sqrt{3}$. Instead of
continuously transforming into a 3$\triangle$ structure, the system then
phase separates into 1$\triangle$ and 2$\square$. This phase separation can be understood in terms of a free energy minimization criterion, equivalent to the double-tangent construction. A similar argument applies to the subsequent transitions
$2\square\to 2\triangle$, $2\triangle\to 3\square$, and so on. Around
7 layers the square symmetric structures eventually disappear, in
favor of the denser hexagonal packings, which have a face-centered cubic (fcc), hexagonal close packed (hcp) or a random hexagonal close packed (rhcp) structure, similar to the crystalline bulk phases of hard spheres.

The sequence of alternating hexagonal and square packings under strong
harmonic confinement corresponds to the simple Pieranski picture of
hard spheres confined between two hard walls
\cite{pieranski}. However, we do not see any of the buckled, rhombic
and prism phases that are found to interpolate between these packings under hard-confinement conditions, when entropy solely determines the phase behavior \cite{pansu,neser,fortini}. By contrast, at high spring constants the behavior of the harmonically confined system is energy-dominated and the usual intervening phases are found to be unstable, because the second derivative of the (free) energy with respect to the density is negative. Furthermore, the intervening phases also appear to be unstable at finite spring constants, as we always observed a spontaneous melting (low spring constants) or phase separation (higher spring constants) of the system when it was initially prepared in one of these phases. In the absence of any stable intervening phases at higher spring constants, the coexistence regions between the stable hexagonal and square packed phases -- which can only exist with low free energy at certain densities due to the integer number of layers -- are wider than under hard-confinement conditions (note, by the way, that in order to approach the hard-confinement limit one would need to increase the exponent of the confining potential, rather than the pre-factor). We point out that the alternating hexagonal and
square-symmetric packings appear to be a common property of different types of confined repulsive particle systems \cite{lowen,kahn,fortini,oguz}, while the character of the intervening phases seems to depend more strongly on the exact details of the particle-particle interactions and the confining potential. 
For example, at high spring
constants we do not observe any new phases between 1$\triangle$ and
2$\square$ in the harmonic potential, while the same hard spheres in a
hard slit would form an intermediate buckled bilayer structure ($2\mathcal{B}$)
\cite{pansu,neser,fortini}, and while charged particles in an
effective harmonic potential are expected to form the sequence
$1\triangle\to 3\triangle\to 2\square$ in the limit $T\to 0$
\cite{oguz}. In the latter system, which considered point-like
particles, the long-ranged repulsive Coulomb interactions between the
particles compete with the attraction to the minimum of the confining
potential.

Remarkably, as the harmonic confinement becomes softer
(lower spring constant), we do find stable intervening phases, which,
however, are not ordered. Instead of the more commonly observed
solid-to-solid transformations, the system undergoes a couple of
re-entrant melting transitions that give rise to intervening liquid
phases, with triple points around \{$k=120$, $\rho=1.2$\} and
\{$k=45$, $\rho=2.25$\}. Thus, around $k\approx 100$ we observe freezing into a stable 1$\triangle$ phase, which then re-melts into a
disordered liquid before freezing again into the 2$\square$ phase (Fig. \ref{fig2}),
while for higher densities we find the same alternating sequence of
hexagonal and square packings as before. We further see that the
softer the confinement, the higher the density at which the initial liquid phase still persists and at sufficiently low spring constants ($k\lesssim 80$) the first stable ordered structures
actually consist of more than one layer. For example, at $k\approx 40$ the first ordered
phase is 2$\triangle$, followed by re-entrant melting, the 3$\square$
structure, and then the other multi-layered ordered phases, whereas
for $k\lesssim 35$ we only observe the latter, without any
intermediate melting.
\begin{figure}[t]
\includegraphics[width=0.45\textwidth]{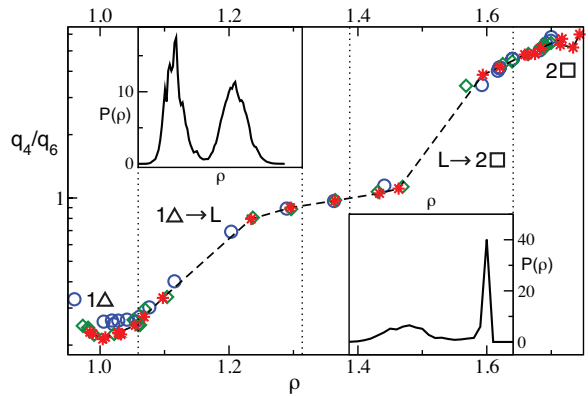}
\caption{\label{fig4} (Color Online) Density-dependent re-entrant
  melting between the 1$\triangle$ and 2$\square$ phases, for $k=100$
  and different sizes of the simulation box ($L=L_x=L_y$). Red stars:
  $L=50$. Green diamonds: $L=20$. Blue circles: $L=10$. The insets
  give the probability distribution of observing a given density at
  coexistence chemical potential on the same density scale as the main
  plot.}
\end{figure}

\begin{figure}[t]
\vskip 2mm
\includegraphics[width=0.45\textwidth]{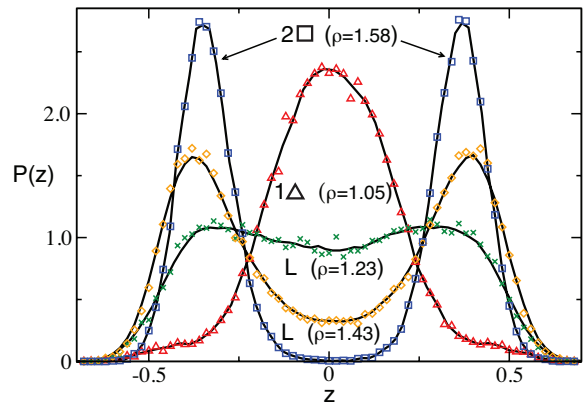}
\caption{\label{fig5} (Color Online) Particle distribution
  along the axis of confinement ($z$), for $k=100$ and different
  densities. Red triangles: 1$\triangle$, $\rho=1.05$. Green crosses:
  liquid, $\rho=1.23$. Yellow diamonds: liquid, $\rho=1.43$. Blue
  squares: 2$\square$, $\rho=1.58$.}
\end{figure}

To make sure that the re-entrant liquid phases did not represent
finite-size artifacts, we performed simulations with three different
box sizes ($L_x=L_y=$ 10, 20 or 50) for $k=100$, Fig. \ref{fig4}. It
can be seen that the results for the different system sizes are
essentially identical and that there is a clear transition from the
1$\triangle$ phase to a re-entrant liquid and then to the 2$\square$
phase, as reflected by the $q_4/q_6$ bond order parameter ratio. We also
determined the density probability distributions at coexistence
\cite{Errington,Martinez} using histogram-reweighing \cite{Ferrenberg}
and expanded ensemble \cite{Lyubartsev} simulations (insets in
Fig. \ref{fig4}). For both the 1$\triangle\to L$ and the
$L\to2\square$ transitions we find bimodal probability distributions that
are characteristic of a first-order phase transition. In addition,
Fig. \ref{fig5} shows the spatial distribution of the particles in the
soft confining potential for different densities outside the
coexistence region. As expected, in the 1$\triangle$ phase the
particle distribution has a single peak centered at the energy minimum
$z=0$ and in the 2$\square$ phase there are two peaks which are
symmetrically located with respect to $z=0$. The intervening
re-entrant liquid starts out with a flat and broad distribution of the
particles at low densities, which then continuously transforms into a
doubly peaked profile as the density increases. Interestingly, this
occurs without further ordering of the particles in the $xy$ plane, so
that the result is a layered liquid. Others have shown that if the same hard spheres are
confined between impenetrable hard walls, a highly ordered buckled
structure forms when the wall separation is larger than required for
the 1$\triangle$ phase, but too small for the 2$\square$ phase
\cite{pansu,neser,fortini}. This buckled $2\mathcal{B}$ phase optimizes the particle
packing in the available space between the two walls by splitting the 1$\triangle$ structure
into rows of particles that alternate in height, effectively forming a
crystal of two interpenetrating rectangular layers, which continuously
transforms into the 2$\square$ structure as the wall separation
increases. In the harmonically constrained system, the re-entrant layered liquid fulfills a similar
interpolating role between the one and two-layer ordered phases, but
the additional energy considerations and greater positional freedom
under soft confinement, together with the absence of long-ranged
particle interactions favor disordered over ordered phases.

\section{Conclusions}
We have identified the stable phases of hard spheres under harmonic
confinement, as a function of the density and the softness of the
confining potential. We find the well-known
`base' sequence of alternating hexagonal and square-symmetric packings
($1\triangle\to 2\square\to 2\triangle\to 3\square\to
3\triangle\to\ldots$), which thus appears to be quite insensitive to
the details of the (confining) interactions. When we look at the presence of stable intervening phases
the soft-confined hard-sphere system is clearly different from hard-confined systems,
though. Instead of the usual highly ordered interpolating particle
packings, we observe phase separation between the hexagonally and
square packed structures under strong confinement, and disordered
re-entrant liquid ($L$) phases under weaker confinement. For
1$\triangle\to L$ and $L\to2\square$, we have shown that the
re-entrant melting/freezing transition has a first-order character and
that the liquid develops a layered structure, without further ordering
of the particles. We argue that the re-entrant and phase separating
behaviors are due to the fact that the penetrable soft harmonic
potential on the one hand offers more positional freedom than
hard-wall confinement does, while, on the other hand, making an
important energy contribution to the overall free energy of the system which depends on the exact particle positions (in
addition to the usual entropic considerations for hard spheres). The
net result is a smaller diversity of crystalline packings, as compared
to hard spheres between two hard walls. We expect that other soft
confining potentials may well have a similar effect, although the
details of the phase diagram will likely be different. The fact that
certain ordered structures can be reliably obtained, while many other
structures can be induced or suppressed on demand through a variation
of the confinement conditions is important for applications that
involve self-assembled structures of colloidal
particles. Interestingly, it should be possible to realize harmonic
confining potentials with dynamically tunable softness experimentally,
for instance in a dielectrophoretic trap.

\begin{acknowledgments}
This work was supported through the research programme of the
Foundation for Fundamental Research on Matter (FOM) which is part of the Netherlands Organisation for
Scientific Research (NWO), by the Slovenian
Research Agency through the Grant P1-0055, by the European Research Council through the Advanced Research Grant COLSTRUCTION (RG52356), and by the 7th Framework Programme through the ITN network COMPLOIDS (RG234810).  DF acknowledges Wolfson Merit Award 2007/R3 of the Royal Society of London  and EPSRC Programme Grant EP/I001352/1. TC acknowledges the support of the Erasmus
work placement scheme.
\end{acknowledgments}


\begin{thebibliography}{}
\bibitem{walz} J.~Walz, I.~Siemers, M.~Schubert, W.~Neuhauser, and
  R.~Blatt, Europhys. Lett. {\bf 21}, 183 (1993)
\bibitem{totsuji} H.~Totsuji, T.~Kishimoto, and C.~Totsuji,
  Phys. Rev. Lett. {\bf 78}, 3113 (1997)
\bibitem{messina} R.~Messina and H.~L\"owen, Phys. Rev. Lett. {\bf 91},
  146101 (2003)
\bibitem{esfarjani} K.~Esfarjani and Y.~Kawazoe, J. Phys.: Condens. Matter
  {\bf 7}, 7217 (1995)
\bibitem{lowen} H.~L\"owen, J. Phys.: Condens. Matter {\bf 21}, 474203
  (2009)
  \bibitem{krishnan} M.~Krishnan, N.~Mojarad, P.~Kukura, and V.~Sandoghdar,
  Nature {\bf 467}, 692 (2010)
  \bibitem{leckband} D.~Leckband and J.~Israelachvili, Quart. Rev. Biophys. {\bf 34}, 105 (2001)  
  \bibitem{copley} A.~L.~Copley~(Ed.) , {\sl Proceedings of the Fourth International Congress on Rheology}, John Wiley \& Sons Inc., New York (1963)
  \bibitem{cohen} I.~Cohen, T.~G.~Mason, and D.~A.~Weitz, Phys. Rev. Lett. {\bf 93}, 046001 (2004)
   \bibitem{vlasov} Y.~A.~Vlasov, X.-Z.~Bo, J.~C.~Sturm, and D.~J.~Norris, Nature {\bf 414}, 289 (2001)
  \bibitem{pieranski} P.~Pieranski, L.~Strzelecki, and B.~Pansu,
  Phys. Rev. Lett. {\bf 50}, 900 (1983)
  \bibitem{pansu} B.~Pansu, Pi.~Pieranski, and Pa.~Pieranski,
  J. Phys. (Paris) {\bf 45}, 331 (1984)
\bibitem{neser} S.~Neser, C.~Bechinger, and P.~Leiderer,
  Phys. Rev. Lett. {\bf 79}, 2348 (1997)
 \bibitem{fortini} A.~Fortini and M.~Dijkstra, J. Phys.: Condens. Matter
  {\bf 18}, L371 (2006)
\bibitem{schmidt} M.~Schmidt and H.~L\"owen, Phys. Rev. Lett. {\bf 76},
  4552 (1996)
    \bibitem{schmidt2} M.~Schmidt and H.~L\"owen,
  Phys. Rev. E {\bf 55}, 7228 (1997)
\bibitem{dubin} D.~H.~E.~Dubin, Phys. Rev. Lett. {\bf 71}, 2753 (1993)
\bibitem{fontecha} A.~B.~Fontecha, H.~J.~Sch\"ope, H.~K\"onig, T.~Palberg, R.~Messina, and H.~L\"owen, J. Phys.: Condens. Matter
  {\bf 17}, S2779 (2005)
\bibitem{maret} K.~Zahn, R.~Lenke, and G.~Maret, Phys. Rev. Lett. {\bf 82}, 2721 (1999)
\bibitem{Osterman} N.~Osterman, D.~Babi\v c, I.~Poberaj, J.~Dobnikar, and P.~Ziherl, Phys. Rev. Lett. {\bf 99}, 248301 (2007)
\bibitem{Dobnikar} J.~Dobnikar, J.~Fornleitner, and G.~Kahl, J. Phys.:
  Condens. Matter {\bf 20}, 494220 (2008)
\bibitem{docoslis} A.~Docoslis and P.~Alexandridis,
  Electrophoresis {\bf 23}, 2174 (2002)  
\bibitem{sullivan} M.~T.~Sullivan, K.~Zhao, A.~D.~Hollingsworth,
  R.~H.~Austin, W.~B.~Russel, and P.~M.~Chaikin, Phys. Rev. Lett. {\bf 96},
  015703 (2006)
\bibitem{leunissen} M.~E.~Leunissen and A.~van~Blaaderen,
  J. Chem. Phys. {\bf 128}, 164509 (2008)
 \bibitem{ackerson} A.~Chowdhury, B.~J.~Ackerson, and N.~A.~Clark, Phys. Rev. Lett. {\bf 55}, 833 (1985)  
\bibitem{jenkins} M.~C.~Jenkins and S.~U.~Egelhaaf, J. Phys.:
  Condens. Matter {\bf 20}, 404220 (2008)
\bibitem{akcora} P.~Akcora, H.~Liu, S.~K.~Kumar, J.~Moll, Y.~Li,
  B.~C.~Benicewicz, L.~S.~Schadler, D.~Acehan, A.~Z.~Panagiotopoulos,
  V.~Pryamitsyn, V.~Ganesan, J.~Ilavsky, P.~Thiyagarajan, R.~H.~Colby, and
  J.~F.~Douglas, Nature Mater. {\bf 8}, 354 (2009)
\bibitem{eiser} L.~Di~Michele, T.~Yanagishima, A.~R.~Brewer, J.~Kotar,
  E.~Eiser, and S.~Fraden, Phys. Rev. Lett. {\bf 107}, 136101 (2011)
\bibitem{Binks} R.~Aveyard, B.~Binks, and J.~Clint, Adv. Colloid Interf. Sci. {\bf 100-102}, 503 (2003)
\bibitem{Clegg} E.~M.~Herzig, K.~A.~White, A.~B.~Schofield,
  W.~C.~K.~Poon, and P.~S.~Clegg, Nature Mater. {\bf 6}, 966 (2007)
 \bibitem{leunissen2} M.~E.~Leunissen, A.~van~Blaaderen, A.~D.~Hollingsworth, M.~T.~Sullivan, and P.~M.~Chaikin, Proc. Nat. Acad. Sci. USA {\bf 104}, 2585 (2007) 
 \bibitem{oettel} F.~Bresme and M.~Oettel, J. Phys.:
  Condens. Matter {\bf 19}, 413101 (2007)
 \bibitem{delGado} L.~Isa, E.~Amstad, K.~Schwenke, E.~Del~Gado, P.~Ilg,
  M.~Kr\"oger, and E.~Reimhult, Soft Matter {\bf 7}, 7663 (2011)
\bibitem{oguz} E.~C.~O\u{g}uz, R.~Messina, and H.~L\"owen, J. Phys.:
  Condens. Matter {\bf 21}, 424110 (2009)
\bibitem{Frenkel} D.~Frenkel and B.~Smit, {\sl Understanding Molecular
  Simulation}, Academic Press, London (2002)
\bibitem{q} P.~J.~Steinhardt, D.~R.~Nelson, and M.~Ronchetti,
  Phys. Rev. B {\bf 28}, 784 (1983)
\bibitem{Velasco} E.~Velasco and L.~Mederos, Phys. Rev. B {\bf 56}, 2432
  (1997)
  \bibitem{Binder} K.~Binder, S.~Sengupta, and P.~Nielaba, J. Phys.:
  Condens. Matter {\bf 14}, 2323 (2002)
    \bibitem{kahn} M.~Kahn, J.-J.~Weis, and G.~Kahl,
  J. Chem. Phys. {\bf 133}, 224504 (2010)
\bibitem{Errington} J.~R.~Errington and A.~Z.~Panagiotopoulos,
  J. Chem. Phys. {\bf 109}, 1093 (1998)
\bibitem{Martinez} F.~J.~Martinez-Veracoechea, B.~Bozorgui, and D.~Frenkel, Soft Matter {\bf 6}, 6136 (2010)
\bibitem{Ferrenberg} A.~M.~Ferrenberg and R.~H.~Swendsen,
  Phys. Rev. Lett. {\bf 61}, 2635 (1988)
\bibitem{Lyubartsev} A.~P.~Lyubartsev, A.~A.~Martsinovski,
  S.~V.~Shevkunov, and P.~N.~Vorontsovvelyaminov, J. Chem. Phys. {\bf 96},
  1776 (1992)
\end{thebibliography}
\end{document}